\documentstyle[prb,aps,twocolumn]{revtex}
\begin{document}
\draft
%@twocolumnfalse\endcsname
\title{ Polarization Dependence of Anomalous X-ray Scattering \\ in Orbital Ordered Manganites}
\author{Sumio Ishihara and Sadamichi Maekawa}
\address{Institute for Materials Research, Tohoku University, 
Sendai, 980-8577 Japan}
\date{\today}
\maketitle
\begin{abstract}
In order to determine types of the orbital ordering in manganites, 
we study theoretically the polarization dependence of the anomalous X-ray scattering 
which is caused by the 
anisotropy of the scattering factor. 
The general formulae of the scattering intensity in 
the experimental optical system is derived and the 
atomic scattering factor is calculated in the microscopic electronic model. 
By using the results, the X-ray scattering intensity in several types of the orbital ordering 
is numerically calculated as a function of azimuthal and analyzer angles. 
\end{abstract}
\pacs{ 71.10.-w, 75.90.+w, 78.70.Ck}
\section{Introduction}
Perovskite manganites and their related compounds, 
$\rm (R_{1-x}A_x)_{1+n} Mn_n O_{3n+1}$  ($n=1,2,\infty$, 
$\rm R=$ rare earth ion, $\rm A=$ alkaline earth ion), 
have been studied extensively 
from the fundamental scientific view point as well as the technological one, 
since the discovery of the colossal magnetoresistance (CMR). \cite{chaha,hel,toku,jin}
In order to explain the unique magnetic and transport properties, 
special attention is paid to the orbital degrees of freedom 
in Mn ions. \cite{good,kana,kugel,hama,mizo,ishi1,ishi2}
The electron configuration of a $\rm Mn^{3+}$ ion is represented by 
$(t_{2g})^3(e_g)^1$ due to the strong Hund coupling. 
Since an electron occupies the two degenerate $e_g$ orbitals, 
the ion has the orbital degrees of freedom as well as 
spin and charge.  
It was supposed that the orbital ordering is realized 
in insulating compounds 
$\rm RMnO_3$ \cite{good,kana,wollan,matsu},  
$\rm R_{0.5}A_{0.5}MnO_3$ \cite{good,wollan,jirak,chen}
and $\rm La_{0.5}Sr_{1.5}MnO_4$. \cite{stern} 
Furthermore, in the metallic phase,  
roles of the orbital degrees in the magnetic, optical 
and transport phenomena were also stressed. \cite{ishi2,maezo,imada,shiba}
However, the direct observation of the orbital degrees was restricted experimentally 
to the case where the polarized neutron diffraction is utilized. \cite{akimitsu}
\par
Recently, Murakami {\it et al.} have applied the anomalous X-ray scattering 
in order to detect the orbital ordering in single layered manganites 
$\rm La_{0.5}Sr_{1.5}MnO_4$. \cite{mura1}
They focused on a reflection at $(\frac{3}{4} \ \frac{3}{4} \ 0)$ point  
and observed a resonant-like peak near the K-edge of a $\rm Mn^{3+}$ ion 
below about 200K. 
They further observed the unique polarization dependence 
which is attributed to the tensor character of the anomalous scattering factor. 
Since all $\rm  Mn^{3+}$ ions are equivalent, 
the reflection at $(\frac{3}{4} \ \frac{3}{4} \ 0)$ is forbidden. 
Therefore, an appearance of the intensity implies that 
two kinds of orbital are alternately 
aligned in the $\rm MnO_2$ plane. 
This type of the orbital ordering is termed the antiferro-type orbital ordering, hereafter. 
The experimental results also 
imply that the dipole transition between Mn $1s$ and Mn $4p$ 
orbitals causes the scattering. 
The experimental method was extended to 
$\rm La_{1-x}Sr_{x}MnO_3$ with $x=0.0$ \cite{mura2}
and $0.12$. \cite{mura3} 
\par
In this paper, 
we study theoretically the polarization dependence of the 
anomalous X-ray scattering 
as a probe to identify types of the orbital ordering 
in manganites. 
Although the phenomenological model for the 
anomalous scattering factor was used to analyze the experimental results  
of $(\frac{3}{4} \ \frac{3}{4} \ 0)$ reflection, 
it was not determined 
which type of the orbital ordering, $(3d_{3x^2-r^2}, 3d_{3y^2-r^2})$
or $(3d_{z^2-x^2}, 3d_{y^2-z^2})$, is realized in $\rm La_{0.5}Sr_{1.5}MnO_4$. \cite{mura1}
As for the theoretical side, 
several types of the orbital ordering is predicted 
in the wide range of the carrier concentration. 
\cite{good,kugel,hama,mizo,ishi1,ishi2,maezo,ishi3,koshi1,koshi2,shii,mizo2}
Once a method to identify types of the orbital ordering is established, 
this becomes applicable to the study of the orbital structure 
in not only manganites but also a wide range of transition metal oxides. 
Considering the fact that types of the orbital ordering 
directly reflects on the 
anisotropy of tensor element of the anomalous scattering factor, \cite{tem,domi}
we investigate the polarization dependence of the scattering intensity. 
Through the numerical calculation in several orbital ordered cases, 
we propose the method to identify the orderings. 
\par
In Sect. II, the general formulae of the polarization dependence of the 
scattering intensity is derived. 
In Sect. III, the anomalous part of the atomic structure factor 
is calculated in a $\rm MnO_6$ cluster. \cite{ishi4} 
In Sect IV, 
the numerical results of the polarization dependence in several orbital ordered cases, 
which are calculated by using the results obtained in Sects. II and III, are presented. 
Sect. V is devoted to the summary and discussion. 
\section{general formulae}
In this section, we derive the general formulae of the polarization 
dependence of the scattering intensity in orbital ordered manganites. 
The conventional experimental arrangement, where the polarization measurement 
is carried out, is shown in Fig.1.\cite{mura1,nagano} 
It mainly consists of a sample crystal and the polarization analyzer 
including analyzer crystal and photon detector. 
The polarization scan is characterized by two rotation angles, i.e., the azimuthal 
angle $(\varphi)$ and the analyzer one ($\varphi_A$). 
The former is the rotation angle of the sample around the scattering vector 
$\vec K=\vec k''- \vec k'$, and the 
latter is that of the analyzer around an axis 
which is parallel to the scattered photon beam. 
The incident beam from the synchrotron source is almost perfectly polarized in the horizontal plane, 
i.e., $\sigma$-polarization. 
The direction of the electric vector of the incident photon with respect to 
the crystalline axis is changed by the azimuthal rotation. 
The scattered photon has both $\pi$- and $\sigma$-polarization components, 
due to the tensor character of the anomalous scattering factor, 
which are separated by the analyzer scan. 
\par
In this optical system, 
the scattering intensity is given by \cite{nagano,kirfel}
\begin{eqnarray}
I(\vec{k'},\vec{k''},\varphi,\varphi_A)=\sum_{\lambda'}
\bigl |
\sum_{\lambda} M_{\lambda' \lambda}(\varphi_A) A_{\lambda \sigma}(\vec{k'},\vec{k''},\varphi)
\bigr |^2 \ ,  
\label{eq:intensity}
\end {eqnarray}
for $\lambda(\lambda')$ =$\sigma$ or $\pi$. 
$M_{\lambda' \lambda}(\varphi_A)$ is the scattering matrix in the analyzer 
and described by the analyzer angle and the scattering one $\theta_A$ 
in the analyzer 
as follows, 
\begin{equation}
 M_{\lambda' \lambda}(\varphi_A)= F_A
\pmatrix{ \cos \varphi_A & -\sin \varphi_A  \nonumber \cr
         \sin \varphi_A \cos 2 \theta_A & \cos \varphi_A \cos 2 \theta_A \cr
}  \ , 
\label{eq:analyser}
\end {equation}
where $F_A$ 
is the scattering factor in the analyzer crystal. 
In this paper, the scattering angle is fixed at $\theta_A=\pi/4$, 
as chosen in the conventional experiments. 
The $\sigma$- and $\pi$-polarized components in the scattered photon are detected 
by the photon detector with $\varphi_A=0$ and $\pi/2$, respectively. 
$A_{\lambda'' \lambda'}(\vec{k''},\vec{k'},\varphi)$ in Eq. (\ref{eq:intensity})
is the scattering amplitude with the incident (scattered) photon momentum 
$\vec{k'}$ $(\vec{k''})$ and the polarization $\lambda'$ ($\lambda''$) as a fuction 
of the azimuthal angle. 
It is given by, 
\begin{eqnarray}
A_{\lambda'' \lambda'}(\vec{k''},\vec{k'},\varphi)&=&\frac{e^2}{mc^2} \sum_{\alpha'' \alpha'}
\epsilon^{(s)}_{ \lambda'' \alpha''}  \nonumber \\
&\times& \biggl [
U(\varphi) V 
F(\vec{k'}, \vec{k''})
 V^\dagger U(\varphi)^\dagger 
\biggr]_{\alpha'' \alpha'}
\epsilon^{(i)t }_{\lambda' \alpha'} \ ,  
\label{eq:amplitude}
\end {eqnarray}
where $F_{\alpha \beta}(\vec k', \vec k'')$ is the structure factor of the sample crystal 
defined in the coordinate of the crystallographic axis 
described as $(\hat a, \hat b, \hat c)$.   
$\epsilon^{(i)}_{ \lambda \alpha}$ and $\epsilon^{(s)}_{ \lambda \alpha}$
are the polarization vectors of the incident and scattered photons, respectively, 
and their explicit forms are given by 
\begin{equation}
\epsilon^{(i)}_{\lambda \alpha}= 
\pmatrix{ 1 & 0 & 0  \cr
         0 & \sin \theta  & \cos \theta }  \ , 
\label{eq:polvec1}
\end {equation} 
and 
\begin{equation}
\epsilon^{(s)}_{\lambda \alpha}= 
\pmatrix{ 1 & 0 & 0  \cr
         0 & -\sin \theta & \cos \theta }  \ , 
\label{eq:polvec2}
\end {equation} 
where the suffix $\alpha$ 
denotes the Cartesian coordinate in the laboratory system, i.e., 
$(\hat e_1, \hat e_2, \hat e_3)$ shown in Fig. 1. 
$\epsilon^{(i)t}_{\lambda \alpha }$ is the transposed matrix of 
$\epsilon^{(i)} _{\lambda \alpha }$.
The unitary matrix $ U(\varphi) $ in Eq. (\ref{eq:amplitude}) describes the azimuthal rotation 
of the sample around the $\hat e_3$-axis defined by  
\begin{equation}
U(\varphi)= 
\pmatrix{ \cos \varphi & -\sin \varphi & 0  \cr
          \sin \varphi &  \cos \varphi & 0  \cr
          0         &  0         & 1   }  \ , 
\label{eq:azimuth}
\end {equation} 
and 
the unitary matrix $V$ describes the transformation from the coordinate 
in the crystallographic axis 
$(\hat{a},\hat{b},\hat{c})$ to that in the laboratory system 
$(\hat{e_1},\hat{e_2},\hat{e_3})$. 
The structure factor 
$F_{\alpha \beta}(\vec{k'},\vec{k''})$ in Eq.(\ref{eq:amplitude})
is represented by a sum of the normal and anomalous parts of the scattering 
factor as follows: 
\begin{equation}
{F}_{\alpha \beta}(\vec{k'},\vec{k''})=N\sum_{i \in cell} 
e^{i(\vec{k'}-\vec{k''}) \cdot \vec{r_i}} f_{i \alpha \beta} (\vec{k'},\vec{k''})  \ , 
\label{eq:structure}
\end {equation} 
with 
\begin{equation}
f_{i \alpha \beta} (\vec{k'},\vec{k''})
=f_{0i} (\vec{k'},\vec{k''}) \delta_{\alpha \beta}   
+\Delta f_{i\alpha \beta}(\vec{k'},\vec{k''}) \ . 
\label{eq:fff}
\end {equation} 
Here, $f_{0i}(\vec{k'},\vec{k''})$ and $\Delta f_{i \alpha \beta}(\vec{k'},\vec{k''})$
are the normal and anomalous parts of the atomic scattering factor, respectively,  
of the $i$-th atom defined in the $(\hat{a},\hat{b},\hat{c})$-coordinate. 
$N$ is the number of the unit cell. 
When we determine the unitary matrix $V$ and 
the atomic scattering factors, 
the scattering intensity is calculated as a function of 
the azimuthal angle $\varphi$  and the analyzer angle $\varphi_A$.
\par
It is instructive to demonstrate the polarization dependence of the scattering intensity 
in the following two cases: 
(i)  $(\hat{e_1},\hat{e_2},\hat{e_3})=(\hat{a},\hat{b},\hat{c})$ 
(azimuthal rotation around the $c$-axis) and 
(ii)  $(\hat{e_1},\hat{e_2},\hat{e_3})=
(\hat{c},\frac{1}{\sqrt{2}}(\hat{a}-\hat{b}),
\frac{1}{\sqrt{2}}( \hat{a}+\hat{b}))$ 
(azimuthal rotation around the $\hat{a}+\hat{b}$-axis). 
When the crystallographic axes coincide with the 
principle ones in the $\rm MnO_6$ octahedron,
the anomalous part of the scattering factor in Eq. (\ref{eq:fff}) is diagonal as 
$\Delta f_{i \alpha \beta}(\vec{k'},\vec{k''})=
\delta_{\alpha \beta}  \Delta f_{i \alpha \alpha}(\vec{k'},\vec{k''})$. 
In this condition, the scattering intensity is given by 
\begin{eqnarray}
I(\vec{k'},\vec{k''},\varphi,\varphi_A)&=& \biggl ( \frac{e^2}{mc^2} \biggr )^2
|F_A|^2  \nonumber \\
&\times &
|D_{\sigma \sigma} \cos \varphi_A-D_{\pi \sigma} \sin \varphi_A|^2 \ , 
\label{eq:iexam}
\end {eqnarray} 
with 
\begin{equation}
D_{\sigma \sigma}=F_{x x} \cos^2 \varphi+F_{y y} \sin^2 \varphi \ , 
\label{eq:d1}
\end {equation} 
and 
\begin{equation}
D_{\pi \sigma}=(-F_{x x}+F_{y y}) \sin \varphi \cos \varphi \sin \theta \ , 
\label{eq:d2}
\end {equation} 
for case (i), and 
\begin{equation}
D_{\sigma \sigma}=F_{z z} \cos^2 \varphi +\frac {1}{2}(F_{x x}+F_{y y}) \sin^2 \varphi \ , 
\label{eq:d3}
\end {equation} 
and 
\begin{eqnarray}
D_{\pi \sigma}&=&\frac{1}{2}(F_{x x}+F_{y y}-2F_{z z}) \sin \varphi \cos \varphi \sin \theta  
\nonumber \\
&+&\frac{1}{2}(-F_{x x}+F_{y y}) \sin \varphi \cos \theta \ , 
\label{eq:d4}
\end {eqnarray} 
for case (ii). 
\par
\section{atomic scattering factor}
The normal and anomalous parts of the 
atomic scattering factor in Eq.(\ref{eq:fff}) 
are obtained by the perturbation calculation with respect to the 
electron-photon interaction. 
The normal part is given by the Fourier transform of the charge density 
$\rho_i$ in the $i$-th atom defined 
by
%\begin{equation}
$f_{0i}=\langle f | \rho_i(\vec K=\vec k''-\vec k') | 0 \rangle \ , $
%\end{equation}
where $| 0 \rangle $ ($| f \rangle $) is the initial (final) electronic state 
with energy $\varepsilon_0$ ($\varepsilon_f$).  
The anomalous one is derived by the interaction between electronic 
current and photon and is expressed as follows: \cite{kolp,blum}
\begin{eqnarray}
\Delta f_{i \alpha \beta}(k',k'') &=& 
  {m \over e^2} \sum_l 
  \Biggl \{ { 
  \langle f | j_{i \alpha}(\vec{-k'}) | l \rangle 
  \langle l | j_{i \beta}(\vec{k''}) | 0 \rangle  \over
\varepsilon_0-\varepsilon_l-\omega_{k''} -i\delta}
\nonumber \\
&+& {
  \langle f | j_{i   \beta}(\vec{ k''} ) | l \rangle 
  \langle l | j_{i \alpha} (\vec{-k'}) | 0 \rangle \over
 \varepsilon_0-\varepsilon_l+\omega_{k'}-i\delta }
\Biggr \} \ \ ,  
\label{eq:fa}
\end {eqnarray}
where $| l \rangle$ is the intermediate electronic state with energy $\varepsilon_l$ 
and $\delta$ is a damping constant. 
The current operator 
%\begin{equation}
$
 j_{i\alpha}(\vec k)={e\over m} \sum_{\sigma}
( A_\alpha(\vec k)  P^\dagger_{i\alpha \sigma} s_{ i \sigma} +h.c.) 
$
%\end{equation}
%
describes the dipole transition between Mn $1s$ and $4p$ orbitals. 
Here, $P_{i \alpha \sigma}$ and $s_{i \sigma}$ 
are the annihilation operators of electron in Mn $4p$ and Mn
$1s$ orbitals, respectively, 
with spin $\sigma$ and Cartesian coordinate $\alpha$. 
$A_\alpha(\vec k)$ is the coupling constant between the current and the photon. 
The contribution from the quadrupole transition is small,  
because the inversion symmetry is preserved in this system. \cite{fab}
As shown in the next section, 
the anisotropy of the tensor element in the anomalous scattering factor 
determines the polarization dependence of the scattering intensity. 
\par
The anomalous part of the 
atomic scattering factor $\Delta f_{i \alpha \beta }(k',k'')$ 
is calculated in the microscopic electronic model. 
We consider a $\rm MnO_6$ octahedron and 
introduce the following orbitals in a Mn ion: 
\{$1s$, $3d_{\gamma}\  (\gamma=\gamma_{\theta+}, \gamma_{\theta-})$, 
$4p_\gamma\  (\gamma=x,y,z)$  \}, 
where 
$| 3d_{\gamma_{\theta +}} \rangle 
=\cos(\theta^{(t)}/2)|3d_{3z^2-r^2} \rangle +\sin(\theta^{(t)}/2)|3d_{x^2-y^2} \rangle$
and 
$| 3d_{\gamma_{\theta -}} \rangle 
=-\sin(\theta^{(t)}/2)|3d_{3z^2-r^2} \rangle +\cos(\theta^{(t)}/2)|3d_{x^2-y^2} \rangle$. 
Six O $2p$ orbitals, which contribute to the $\sigma$-bond with the Mn $3d$ orbitals, 
are also considered. 
These are denoted as 
$2p_{i}$ $(i=1 \sim 6)$ where 
$2p_{1(2)}$, $2p_{3(4)}$ and $2p_{5(6)}$ are $2p_x$ orbital at the position of $(+(-)a/2,0,0)$, 
$2p_y$ at $(0,+(-)a/2,0)$,
and 
$2p_z$ at $(0,0,+(-)a/2)$, respectively, with Mn-O bond length $a/2$. 
The position of the Mn ion is chosen to be $(0,0,0)$. 
By utilizing the irreducible representation in $O_h$ group, 
these O $2p$ orbitals are recombined 
as $ \{ 2p_{\gamma_{\theta+}}, 2p_{\gamma_{\theta-}}, 
2p_{x}, 2p_{y}, 2p_{z}, 2p_{r^2} \}$. 
For example, 
$2p_{3z^2-r^2}$ and $2p_x$ orbitals are represented as
%
%\begin{equation}
$
 2p_{3z^2-r^2}=\frac{1}{\sqrt{3}}(\frac{1}{2}(-2p_1+2p_2-2p_3+2p_4) +2p_5-2p_6)\ , 
$
%\end{equation}
%
and 
$
 2p_{x}=\frac{1}{\sqrt{2}}(2p_1+2p_2) \ .  
$
\par
On these bases, 
we set up the following Hamiltonian:  
\begin{eqnarray}
H=H_0+H_t+H_{3d-4p}+H_{4p-2p}+H_{core}+H_{3d-3d} \ . 
\label{eq:hamiltonian}
\end{eqnarray}
The first and second terms describe the 
energy level in each orbital and the electron transfer between 
Mn $3d$ and O $2p$ orbitals, respectively, 
and are given by 
\begin{eqnarray}
H_0+H_t&=&\sum_{\gamma_\theta \sigma} \varepsilon_d 
d^\dagger_{\gamma_\theta \sigma} d_{\gamma_\theta \sigma}
+\sum_{\gamma \sigma} \varepsilon_P P^\dagger_{\gamma \sigma} P_{\gamma \sigma} 
 \nonumber \\
&+&\sum_{\Gamma=\gamma, \gamma_\theta \ \sigma} 
\varepsilon_p p^\dagger_{\Gamma \sigma} p_{\Gamma \sigma}
+\sum_{\sigma} \varepsilon_s s^\dagger_{\sigma} s_{\sigma}  \nonumber \\ 
&+&\sqrt{3}t_{3d-2p} 
\sum_{ \gamma_{\theta} \sigma } 
\biggl (      d_{\gamma_\theta \sigma}^{\dagger} 
p_{\gamma_\theta \sigma}  +h.c.  \biggr ) \ ,  
\label{eq:h01}
\end{eqnarray}
where, $d_{\gamma_\theta \sigma}$, and $p_{\Gamma \sigma}$
are the electron annihilation operators in 
Mn $3d$, and O $2p$ orbitals, respectively, with 
spin $\sigma$ and orbital $\gamma_{\theta}(=\gamma_{\theta+},\gamma_{\theta-})$
and $\gamma(=x,y,z)$. 
The third and fourth terms in Eq.(\ref{eq:hamiltonian}), 
which describe the Coulomb interaction between Mn $3d$ and Mn $4p$ orbitals and 
Mn $4p$ and O $2p$ orbitals, respectively, 
are given by 
\begin{eqnarray}
H_{3d-4p}+H_{4p-2p}&=&\sum_{\gamma_{\theta} \gamma} 
V(3d_{\gamma_\theta},4p_{\gamma})
\ n(3d_{\gamma_\theta}) \ n(4p_{\gamma})  \nonumber \\
&+&\sum_{\gamma_{\theta} \gamma} 
V(4p_{\gamma},2p_{\gamma_\theta}) \ 
n(4p_{\gamma}) \ n_h(2p_{\gamma_\theta})  \ . 
\label{eq:v3d4p2p}
\end{eqnarray}
$n(3d_{\gamma_\theta})$ and $n(4p_{\gamma})$ are the 
number operators of Mn $3d$ and Mn $4p$ electrons, respectively, and 
$n_h(2p_{\gamma})$ is the number operator of the O $2p$ holes . 
The explicit forms of the Coulomb interaction in Eq.(\ref{eq:v3d4p2p}) 
are written by 
\begin{eqnarray}
V(3d_{\gamma_{\theta \pm}},4p_\gamma)&=& 
F_0(3d,4p)  \nonumber \\
& \pm & 
4F_2 (3d,4p)\cos \Bigl(\theta^{(t)}+m_\gamma {2 \pi \over 3} \Bigr) \ , 
\label{eq:intra}
\end{eqnarray}
and 
\begin{equation}
V(2p_{\gamma_{\theta\pm}},4p_\gamma)= 
-\varepsilon \pm { \varepsilon \rho^2 \over 5} \cos \Bigl(\theta^{(t)}+m_\gamma {2 \pi \over 3} \Bigr) \ . 
\end{equation}
$m_x=+1,$ $m_y=-1,$ and $m_z=0$.  
$F_{n}(3d,4p)$ is the Slater-integral between $3d$ and $4p$ electrons, 
and is defined by 
$F_0(3d,4p)=F^{(0)}(3d,4p)$ and $F_2(3d,4p)={1 \over 35}F^{(2)}(3d,4p)$. 
$\varepsilon=Ze^2/a$ and $\rho=\langle r_{4p} \rangle /a$
where $Z=2$, and $\langle r_{4p} \rangle$ 
is the average radius of Mn $4p$ orbital. 
Furthermore, we introduce the interaction between Mn $3d(4p)$ electron and Mn $1s$ core hole  as follows; 
\begin{eqnarray}
H_{core}&=&\sum_{\gamma_\theta \sigma \sigma'} 
V(1s,3d) \ n(3d_{\gamma_\theta \sigma}) \  n(1s_{\sigma'})  \nonumber \\
&+&\sum_{\gamma \sigma \sigma'} V(1s,4p) \ n(4p_{\gamma \sigma}) \  n(1s_{\sigma'}) \ , 
\end{eqnarray}
where 
both $V(1s,3d)$ and $V(1s,4p)$ 
do not depend on the orbitals. 
The last term in Eq.(\ref{eq:hamiltonian}), $H_{3d-3d}$,  
is the interaction between Mn $3d$ electrons 
\cite{ishi1} defined by 
\begin{eqnarray}
H_{3d-3d}&=&U\sum_{\gamma_\theta} \ 
    n(3d_{\gamma_\theta \uparrow}) \  n(3d_{\gamma_\theta \downarrow})  \nonumber \\
&+&U'  n(3d_{\gamma_{\theta+}}) \  n(3d_{\gamma_{\theta-}}) \nonumber \\
\\
&+&K\sum_{\sigma \sigma'} 
d_{\gamma_{\theta+}  \sigma }^\dagger  d_{\gamma_{\theta-}  \sigma'}^\dagger
d_{\gamma_{\theta+}  \sigma'}           d_{\gamma_{\theta-}  \sigma}  \nonumber \\
&-&J_H \sum_{\gamma_\theta} \vec S_{t_{2g}} \cdot \vec s_{\gamma_\theta}  \ , 
\label{eq:h3d3d}
\end{eqnarray}
where $U$, $U'$ and $K$ are 
the Coulomb and exchange interactions between $3d$ electrons with 
$K=(U-U')/2$. 
$J_H$ describes the Hund coupling between the localized $t_{2g}$ 
spin ($ \vec S_{t_{2g}}$) with $S=3/2$ 
and $e_g$ spin defined by 
$\vec s_{\gamma_\theta}=\frac{1}{2} \sum_{\sigma' \sigma''} d_{\gamma_\theta \sigma'}^\dagger 
(\vec \sigma)_{\sigma' \sigma''} d_{\gamma_\theta \sigma''}$. 
The electron transfer ($t_{4p-2p}$) between Mn $4p$ and O $2p$ is not included in the 
model. 
As mentioned above, 
the O $2p$ orbitals are recombined by utilizing the irreducible representation 
in $O_h$ group, and 
the Mn $3d$ and Mn $4p$ orbitals are decoupled when $t_{3d-2p}$ and 
$t_{4p-2p}$ are taken into account. 
Therefore, the essential conclusion about the anisotropy of the atomic 
scattering factor being based on the Coulomb interactions is not changed, 
although a large overlap between Mn $4p$ and O $2p$ orbitals 
may reduce the Coulomb interaction between Mn $3d$ and Mn $4p$ 
and increase the one between Mn $4p$ and O $2p$. 
\par
It is worth noting that 
$H_{3d-4p}$ and $H_{4p-2p}$ provide the 
anisotropy of the atomic scattering factor in orbital 
ordered states. 
The essential points in the mechanism 
of the anisotropy is the fact that 
these interactions 
depend on the occupied Mn $3d$ orbitals. 
When an electron occupies $3d_{3z^2-r^2}$ orbital $(\theta^{(t)}=0)$, 
the energy of $4p_{z}$ orbital is higher than that of $4p_{x(y)}$ orbital 
by $6F_2(3d,4p)$ due to $V(3d_{\gamma_\theta}, 4p_\gamma)$, 
as a result, 
the scattering intensity near the K-edge is dominated by $4p_{x(y)}$ 
orbital. 
Furthermore, 
$ | 3d_{3z^2-r^2}^1 \rangle $ state is strongly 
mixed with $ |3d_{3z^2-r^2}^1 3d_{x^2-y^2}^1 \underline{p_{x^2-y^2}} \rangle $, 
where $ |\underline{p_{x^2-y^2}} \rangle $ 
describes the state where a hole occupies the 
O $ 2p_{x^2-y^2} $ orbital. 
Through the inter-atomic Coulomb interaction $V(2p_{\gamma_\theta}, 4p_\gamma)$,
the energy of $4p_z$ orbital becomes higher than that of $4p_{x(y)}$ orbital by 
$\frac{3}{10} \varepsilon \rho$. 
Therefore, the two interactions due to the intra- and inter-site Coulomb interactions 
result cooperatively in the anisotropy of the scattering factor. 
In the following numerical calculation, 
$H_{4p-2p}$ is often neglected, 
since the inter-site interaction is well screened in comparison with  
the intra-site Coulomb interaction. 
\par
The lattice distortion in the $\rm MnO_6$ octahedron also 
becomes the origin of the anisotropy of the 
scattering factor. 
When there is the Jahn-Teller(JT)-type 
lattice distortion characterized by the difference of the bond lengths 
in the $ab$-plane and the $c$-direction, i.e., $\delta a=a_z-a_{x(y)}$, 
the energy difference of the Mn $4p_{x(y)}$ and $4p_z$ is brought about. 
In the case of $a_z > a_{x(y)}$, 
the energy of the Mn $4p_{z}$ orbital is relatively stabilized in comparison with 
that of Mn $4p_{x(y)}$ orbital. 
Therefore, 
this contribution competes with the above originated from the intra- and inter-atomic 
Coulomb interactions. 
In $\rm LaMnO_3$ where the $\delta a/a$ is about $15\%$, \cite{kawa}
the contributions from the Coulomb interactions seems to be weakened by 
that from the lattice distortion. 
However, 
the notable lattice distortion in $\rm MnO_6$ octahedron is not observed in 
$\rm La_{0.5}Sr_{1.5}MnO_4$. \cite{mura4}
Furthermore, 
it was recently reported that 
in $\rm La_{0.88}Sr_{0.12}MnO_3$ 
the X-ray scattering intensity attributed to the anomalous scattering  
is observed at (030) reflection in the $O^{\ast}$ phase, \cite{mura3}
where the six Mn-O bonds in the $\rm MnO_6$ octahedron are equivalent. \cite{kawa}
Therefore, we conclude that 
the Coulomb interactions dominate the origin of the anisotropy of the anomalous scattering factor 
in these compounds
and do not introduce the 
contribution from the lattice distortion in the Hamiltonian in Eq.(\ref{eq:hamiltonian}). 
\par
By adopting the above Hamiltonian, 
we calculate the atomic scattering factor in 
the configuration interaction method. 
$| 3d_{\gamma_{\theta+}}^1 \rangle$ and 
$| 3d_{\gamma_{\theta+}}^1  3d_{\gamma_{\theta-}}^1  
 \underline{2p_{\gamma_{\theta-}}}  \rangle$
states are introduced as the bases of the initial and final states, 
and  
$| 3d_{\gamma_{\theta+}}^1 4p_{\gamma}^1 \ \underline{1s} \rangle$ and 
$| 3d_{\gamma_{\theta+}}^1  3d_{\gamma_{\theta-}}^1  
 \underline{2p_{\gamma_{\theta-}}}  4p_{\gamma}^1 \ \underline{1s} \rangle$
states are as the bases of the intermediate states. 
The second term in the right hand side in Eq. (\ref{eq:fa}) is 
only considered. 
The parameter values in the Hamiltonian 
are chosen to be 
$\varepsilon_P-\varepsilon_d=13.0eV$, 
$\varepsilon_p-\varepsilon_d=2.5eV$, 
$t_{3d-2p}=1.2eV$, $V(3d,1s)=V(4p,1s)=7.0eV$, $U=8.5eV$, $U'=6.5eV$ 
and $J_H=1.0eV$. 
The Slater-integrals between Mn $4p$ and Mn $3d$ electrons are chosen to be 
$F_0(3d,4p)=5.0eV$ and $F_2(3d,4p)=0.3eV$ 
which are the same order as ones 
between Mn $3d$ electrons $F_{n}(3d,3d)$
evaluated by the photoemission experiments. \cite{saitoh}
The damping parameter $\delta$ in Eq. (\ref{eq:fa}) is taken to be 0.5eV. 
\par
In Fig. 2, we present the energy dependence of the 
real and imaginary parts of the anomalous scattering factor. 
The edge of the lowest peak of $\Delta f_{i \alpha \alpha}''$
corresponds to the Mn K-edge.  
Near the edge in Fig. 2(a), where $3d_{3z^2-r^2}$ orbital is occupied, 
the anisotropy is clearly shown. 
The energy position of $\Delta f_{i xx(yy)}''$ is lower and its weight is larger than 
those of $\Delta f_{ i zz}''$.  
In Fig. 2(b) where $3d_{x^2-y^2}$ orbital is occupied, 
the anisotropy in the main peak is entirely opposite to that in Fig. 2(a). 
In both Figs. 2 (a) and (b), 
the spectra of $\Delta f_{i xx}$ and $\Delta f_{i yy}$ are identical 
as expected. 
\section{polarization dependence of the scattering intensity}
By using the general formulae of the scattering intensity (Eq.(\ref{eq:intensity})-(\ref{eq:fff})) 
and the atomic scattering factor calculated in the microscopic electronic structure,  
we numerically calculate the polarization dependence of the 
normalized scattering intensity 
$ \tilde I(\varphi,\varphi_A)
=I(\varphi,\varphi_A) /(N_A^2 e^4/(mc^2)^2|F_A|^2)$, 
where $N_A$ is the number of the Mn atom. 
The energy of the incident photon beam 
is fixed at that of the K-edge in a $\rm Mn^{3+}$ ion. 
In our model, it is chosen to be the energy at which 
the lowest component of $\Delta f''_{ i  \alpha \alpha}$ 
$(\alpha=x,y,z)$
has a half value of its maximum. 
The normal part of the scattering factor 
of a $\rm Mn^{3+}$ ion is chosen as 
$f_{0 i }m/|A_{\alpha \alpha}|^2=5 eV^{-1}$ which 
is about 5 times larger than the typical value of  
$|\Delta f_{ i \alpha \alpha}|m/|A_{\alpha \alpha}|^2$ 
at the edge. \cite{mura1}
The following part in this section is divided into three subsections 
and the numerical results calculated in several orbital ordered cases are presented. 
The qualitative features in the following numerical results, 
e.g. periodicity and phase of the oscillations as a function of $\varphi$, 
are independent of details of the microscopic calculation shown in the previous section 
and are related to symmetry of the orbital ordering and the experimental arrangement, 
although the detail quantitative ones 
depend on the adopted interactions and parameter values in the microscopic model. 
\medskip
\par
\subsection{ ferro-type orbital ordering in the simple cubic lattice} 
We consider the ferro-type orbital ordering (orbital-F) in the simple cubic lattice  
where a kind of orbital is occupied in each Mn site 
and demonstrate how to identify types of orbital through the polarization analyses. 
The numerical results shown in this subsection will help us to understand 
the polarization dependence of the fundamental reflections in the 
antiferro-type orbital ordering shown in the next subsection. 
\par
In orbital-F case, the structure factor per unit cell (Eq. (\ref{eq:structure})) is given by 
%
%\begin{equation}
$\frac{1}{N} F_{\alpha \beta}=f_{A \alpha \beta} $
%\label{eq:ffund1}
%\end {equation} 
%
at the $(h k l)=(n_x  n_y  n_z)$ reflection with integer $n_i (i=x,y,z) $. 
Here, 
$(h k l)$ is the indices in the cubic coordinate defined as 
$\vec K=h \frac{2 \pi}{a} \hat a+k \frac{2 \pi}{a} \hat b+l \frac{2 \pi}{a} \hat c$ 
where $a$ is the lattice parameter in the cubic cell. 
$f_{A \alpha \beta}$ is the sum of the normal and anomalous parts 
in a $\rm Mn^{3+}$ ion with orbital $A$; 
$\delta_{\alpha \beta} f_{0A}+\Delta f_{A \alpha \beta}$. 
In Fig. 3(a), we present the scattering intensity 
with $\theta^{(t)}_A=0$ which means that an electron 
occupies the $3d_{3z^2-r^2}$ orbital in each $\rm Mn^{3+}$ ion. 
It is calculated in the cases of 
$(\hat{e_1},\hat{e_2},\hat{e_3})=(\hat{b},\hat{c},\hat{a})$ 
(azimuthal scan around $a$-axis). 
In the case of $(\hat{e_1},\hat{e_2},\hat{e_3})=(\hat{c},\hat{a},\hat{b})$ 
(azimuthal scan around $b$-axis), 
$\varphi$ dependence is entirely opposite to that in Fig. 3(a).  
On the contrary, 
the intensity is independent of $\varphi$ in the case where 
the azimuthal scan is performed around $c$-axis. 
We first consider the contribution from the normal part of the 
scattering factor. 
The polarization dependence of the intensity is represented by 
$\tilde{I} (\varphi,\varphi_A)=|f_{0A} \cos \varphi_A|^2$, 
which is independent of $\varphi$ due to the scalar character of  $f_{0A}$. 
For convenience, 
the value of $\varphi_A=0$ is defined 
$I_0=|f_{0A}|^2$. 
This $\varphi_A$ dependence  
is interpreted that 
the scattering at $\varphi_A=0$ ($\pi/2$) 
is accompanied with the polarization of 
$(\lambda',\lambda'')=(\sigma,\sigma)$($(\sigma, \pi)$), 
where $\lambda'(\lambda'')$ describes the polarization of the incident(scattered ) photon.  
Therefore, 
the polarization dependence in Fig. 3(a) is dominated 
by the normal part of the scattering factor and slightly modified 
by the anomalous part. 
\par
Let us examine the scattering intensity at $\varphi_A=0$ in Fig. 3(a). 
This is represented by 
\begin{eqnarray}
\tilde {I} (\varphi,\varphi_A=0)=|f_{0A}+\Delta f|^2 \ ,
\label{eq:ffund4} 
\end {eqnarray} 
with 
\begin{eqnarray}
\Delta f=\Delta f_{Ayy} \cos^2 \varphi+\Delta f_{Azz} \sin^2 \varphi   \ . 
\label{eq:ferro}
\end {eqnarray}
Since the normal part of the scattering factor dominates the scattering intensity,  
it is approximated as 
$\tilde {I} (\varphi,\varphi_A=0) \sim f_{0A}^2+2 f_{0A} \Delta f'_{A}$, 
where $\Delta f'_{A}$ is negative at the edge. 
As a result, 
the real part of the anomalous component decreases the scattering intensity from 
$I_0$. 
In the case where $3d_{3z^2-r^2}$ orbital is occupied, 
$|\Delta f'_{Axx(yy)} |$ has a larger value at the edge 
in comparison with $|\Delta f'_{Azz}|$ (see Figs. 2(a) and (c)). 
Therefore, 
the reduction of the scattering intensity from $I_{0}$ is remarkable 
around $\varphi=0$ and $\pi$, where 
the electric vector in the incident photon is parallel to the $b$-axis. 
\par
In Fig. 3(b), we present the polarization dependence of the scattering intensity 
in several orbital-F cases at $\varphi_A=0$. 
With increasing $\theta^{(t)}_A$ from $\theta^{(t)}_A=0$, 
the $\varphi$ dependence becomes weak and disappears at $\theta^{(t)}_A=2\pi/6$ 
($3d_{y^2-z^2}$). 
With further increasing $\theta^{(t)}_A$, 
the phase of a modulation becomes inverse, since $\Delta f_{Azz}$ 
dominates the scattering factor near the edge.  
The inversion of a modulation in $\tilde{I}(\varphi,\varphi_A=0)$ 
occurs at $\theta^{(t)}_A=4\pi/6$ and 
$0$ in the cases of 
$(\hat{e_1},\hat{e_2},\hat{e_3})=(\hat{c},\hat{a},\hat{b})$ 
and 
$(\hat{a},\hat{b},\hat{c})$, respectively. 
By combining the azimuthal scan around $a$-, $b$- and $c$-axes, 
it is possible to identify the occupied orbital in the orbital-F case. 
\par
\subsection{antiferro-type orbital ordering in the simple cubic lattice}
Next, we consider the antiferro-type orbital ordering (orbital-AF),  
where two kinds of orbital sublattice, denoted by $A$ and $B$, exist.
In the simple cubic lattice, 
there are three kinds of orbital-AF, that is, 
layer-type (orbital-AAF), rod-type (orbital-CAF) and NaCl-type (orbital-GAF). 
These notations are defined by analogy with types of antiferromagnetic 
structures. \cite{maezo,ishi3}
In each case, 
two kinds of the reflection point exist, 
termed the fundamental reflection at $(h k l)=(n_x n_y n_z)$ 
and the orbital superlattice reflection at 
$(h k l)=$
$(n_x \  n_y \  n_z+\frac{1}{2})$ for orbital-CAF, 
$( n_x +\frac{1}{2} \  n_y +\frac{1}{2} \  n_z)$ for orbital-AAF and 
$(n_x+\frac{1}{2} \ n_y+\frac{1}{2} \  n_z+\frac{1}{2})$ for orbital-GAF.
By examining the superlattice reflection, 
the above three types of the orbital-AF is able to be distinguished. 
The structure factor per unit cell is written as 
$ \frac{1}{N}F_{\alpha \beta}=(f_{A \alpha \beta}+f_{B \alpha \beta}) $
for the fundamental reflection and  
$ \frac{1}{N}F_{\alpha \beta} = (f_{A \alpha \beta}-f_{B \alpha \beta})$ for 
the superlattice reflection. 
For convenience, 
we define 
$f_{+ \alpha \beta}=\frac{1}{2}(f_{A \alpha \beta}+f_{B \alpha \beta})$ and 
$ f_{- \alpha \beta}=\frac{1}{2} (f_{A \alpha \beta}-f_{B \alpha \beta})$, 
and term the former and latter the ferro- and antiferro-components of the scattering 
factor, respectively.  
We neglect the difference of the normal part of the scattering factors 
of $\rm Mn^{3+}$ ions between the $A$ and $B$ sublattices. 
Therefore, at the orbital superlattice reflection, 
only the antiferro-component of the anomalous part contributes. 
\par
In Fig. 4(a), the polarization dependence of the fundamental reflection in 
the orbital-AF case is presented. 
The orbital state in the two sublattices is $(\theta^{(t)}_A,\theta^{(t)}_B)=(4\pi/6,-4\pi/6)$ 
which corresponds to the ($3d_{3y^2-r^2},3d_{3x^2-r^2}$)-type orbital ordering. 
The coordinate in the laboratory system is chosen to be 
$(\hat e_1, \hat e_2, \hat e_3)=(\hat b, \hat c, \hat a)$. 
The explicit formula of the intensity in Fig. 4(a) at $\varphi_A=0$ is given by 
Eq. (\ref{eq:ffund4}) with
\begin{eqnarray}
\Delta f=\Delta f_{+yy}\cos^2 \varphi +\Delta f_{+zz}\sin^2 \varphi \ . 
\label{eq:anti}
\end{eqnarray} 
We note that 
$\Delta f_{Ayy}$ and $\Delta f_{Azz}$ in Eq.(\ref{eq:ferro})
are replaced by the ferro-components of the scattering factor, i.e., 
$\Delta f_{+yy}$ and $\Delta f_{+zz}$ in Eq.(\ref{eq:anti}), respectively. 
The reduction of the intensity from $I_0$
becomes remarkable around 
$\varphi=\pi/2$ and $3\pi/2$  
where the electric vector in the incident beam is parallel  
to the $c$-axis. 
This is because 
the ferro-component of the scattering factor 
$\frac{1}{2}(\Delta f_{A \alpha \alpha}+\Delta f_{B \alpha \alpha})$ 
is larger
for $\alpha=z$ in comparison with that for $\alpha=x(y)$, 
since the $3d_{3y^2-r^2}$ and $3d_{3x^2-r^2}$ orbitals 
are almost elongated in the $ab$-plane.  
When the azimuthal scan is performed around $c$-axis, 
the intensity is almost independent of $\varphi$, because the 
ferro-component of the scattering factor along the 
$a$- and $b$-axes is identical under the condition of 
$\theta^{(t)}_B= -\theta^{(t)}_A$.
\par
In Fig. 4(b), the orbital 
dependence of the intensity at the fundamental reflection 
with $\varphi_A=0$ is shown. 
With increasing $\theta_A^{(t)}$ from $\theta_A^{(t)}=\pi/6$, 
the $\varphi$ dependence becomes weak and 
is almost smeared out at $(\theta_A^{(t)},\theta_B^{(t)})=(\pi/2,-\pi/2)$. 
With further increasing $\theta_A^{(t)}$, 
the phase of the modulation is reversed. 
In the case of $\theta_A^{(t)} > \pi/2$, 
the ferro-component of the scattering factor in the $c$-direction
becomes dominant in comparison with that in the $ab$-plane. 
We conclude that 
through the measurement of the phase of the polarization dependence, 
we can determine whether $\theta_A^{(t)}$ is larger or smaller than $\pi/2$. 
\par
In Fig. 5(a), the orbital superlattice reflection, 
which appears only in orbital-AF ordering is shown. 
The orbital state is 
$(\theta^{(t)}_A,\theta^{(t)}_B)=(4\pi/6,-4\pi/6)$ 
which corresponds to 
$(3d_{3y^2-r^2},3d_{3x^2-r^2})$ ordering.
The coordinate in the laboratory system is chosen to be 
$(\hat{e_1},\hat{e_2},\hat{e_3})=
(\hat{c},\frac{1} {\sqrt{2}}(\hat a-\hat b),
        \frac{1} {\sqrt{2}}(\hat a+\hat b) )$.   
The azimuthal scan in this coordinate was carried out in 
$\rm La_{0.5}Sr_{0.5}MnO_4$ and $\rm LaMnO_3$. \cite{mura1,mura2}
The explicit formulae of the polarization dependence 
at $\varphi_A=0$ and $\pi/2$  
are given by  
\begin{eqnarray}
\tilde {I} (\varphi,\varphi_A=0)&=& |\Delta f_{-zz}  \cos^2 \varphi \nonumber \\
 &+&\frac{1}{2} \bigl( \Delta f_{-xx}+\Delta f_{-yy} \bigr) \sin^2 \varphi  |^2  \ ,   
\label{eq:forbid10}
\end{eqnarray}
and 
\begin{eqnarray}
\tilde {I} (\varphi,\varphi_A = \pi/2)&=&
| \frac{1}{2} \bigl( -\Delta f_{-xx}+\Delta f_{-yy} \bigr) \sin \varphi \cos \theta \nonumber \\
& + & \frac{1}{2} \bigl(\Delta f_{-xx}+\Delta f_{-yy}-2\Delta f_{-zz}  \bigr)  \nonumber \\
&\times & \sin \varphi \cos \varphi \sin \theta |^2  \ ,   
\label{eq:forbid11}
\end{eqnarray}
respectively. 
The intensity is expressed only by 
the antiferro-component of the scattering factor. 
At this reflection, the information of the orbital ordering is derived 
without disturbance of the normal part of the scattering factor, 
in contrast with the case of the fundamental reflection. 
When the orbital ordering is assumed to be $(\theta^{(t)}_A, \theta^{(t)}_B=-\theta^{(t)}_A)$, 
the condition, 
$\Delta f_{Axx}=\Delta f_{Byy}$, 
$\Delta f_{Ayy}=\Delta f_{Bxx}$ and 
$\Delta f_{Azz}=\Delta f_{Bzz}$,  
is satisfied. 
In this case, 
$\tilde {I} (\varphi,\varphi_A =0)$ in Eq.(\ref{eq:forbid10}) becomes zero, 
and $\tilde {I} (\varphi,\varphi_A =\pi/2)=
|\frac{1}{2} \bigl ( \Delta f_{Axx}-\Delta f_{Ayy} \bigr) 
\sin \varphi \cos \theta|^2$ in Eq. (\ref{eq:forbid11}). 
These expression gives a square of the sinusoidal symmetry in 
the polarization dependence as shown in Fig. 5(a). 
By utilizing this characteristic feature of the polarization dependence, 
we can judge whether the condition 
$(\theta^{(t)}_A, \theta^{(t)}_B=-\theta^{(t)}_A)$ 
is satisfied or not. 
For example, 
as shown in Fig. 6, when we assume different types of the orbital-AF ordering 
$(\theta^{(t)}_A, \theta^{(t)}_B=\theta^{(t)}_A+\pi)$,  
the polarization dependence shows a quite different feature from that in Fig. 5. 
\par
In Fig. 5(b), the polarization dependence at $\varphi_A=\pi/2$ 
in several orbital-AF cases is presented. 
Here, the condition $(\theta^{(t)}_A, \theta^{(t)}_B=-\theta^{(t)}_A)$ 
is assumed. 
The scattering intensity changes with $\theta_A^{(t)}$ and 
becomes maximum around $\theta_A^{(t)}=\pi/2$. 
The polarization dependence is, however, given by a square of the sinusoidal function in all cases. 
In the azimuthal analyses performed in $\rm La_{0.5}Sr_{1.5}MnO_4$ \cite{mura1}, 
the experimental data are well fitted by the above function. 
Therefore, 
we conclude that the orbital ordering in $\rm La_{0.5}Sr_{1.5}MnO_4$ 
satisfies the condition $(\theta^{(t)}_A, \theta^{(t)}_B=-\theta^{(t)}_A)$, i.e., 
the $A$ and $B$ orbitals are symmetric with respect to 
replacement of $x(y)$ by $y(x)$. 
However, we can not determine which type of the orbital ordering, $(3d_{3x^2-r^2}, 3d_{3y^2-r^2})$
or $(3d_{z^2-x^2}, 3d_{y^2-z^2})$ is realized, through the measurement of this reflection. 
\medskip
\par
\subsection {orbital ordering in charge ordered state} 
The charge ordering, \cite{chen,stern,mura1} 
where $\rm Mn^{3+}$ and $\rm Mn^{4+}$ ions are alternately aligned 
in the $\rm MnO_2$ plane, 
have been observed in $\rm R_{1-x}A_{x}MnO_3$ 
near $x=0.5$ and 
$\rm La_{0.5}Sr_{1.5}MnO_4$. \cite{chen,stern,mura1} 
In the latter compound, it is accompanied with 
the orbital ordering \cite{mura1} and the CE-type spin ordering. 
In this subsection, we investigate 
the polarization dependence of the scattering intensity 
in the charge ordered phase and propose 
the possibility to identify the orbital ordering in this phase.  
\par
We consider the alternating alignment of $\rm Mn^{3+}$ and $\rm Mn ^{4+}$ ions 
in the $\rm MnO_2$ plane and 
introduce two kinds of the orbital sublattice for $\rm Mn^{3+}$ ion, 
denoted by $A$ and $B$. 
It is assumed that a $\rm Mn^{3+}$ ion has $A(B)$ orbital,  
when the spin in a $\rm Mn^{3+}$ ion is parallel to 
that in its neighboring $ \rm Mn^{4+} $ ions in the 
$a(b)$-direction. \cite{good,kana,wollan,chen,mura1}
This type of orbital ordering is consistent with the CE-type spin structure.  
Here, we pay our attention to the reflection at 
$(h \ k \ l)=(n_x+\frac{1}{2} \ n_y+\frac{1}{2} \ n_z)$
termed the charge-order reflection, which appears due to the existence of the charge ordering. 
The structure factor at the reflection is given by 
\begin{equation}
F_{\alpha \beta}=
N(\widetilde f_{A \alpha \beta}+ \widetilde f_{B \alpha \beta }-2  f_{4 \alpha \beta })  \ , 
\end {equation} 
where $\widetilde f_{A(B) \alpha \beta}$ 
is the scattering factor defined in the orthohombic coordinate  
as follows 
\begin{eqnarray}
 \widetilde f_{l  \alpha \beta}= 
\pmatrix {
\frac{1}{2} (f_{l xx}+f_{lyy}) &
\frac{1}{2} (f_{l xx}-f_{lyy}) &
0 \cr 
\frac{1}{2} (f_{l xx}-f_{lyy}) &
\frac{1}{2} (f_{l xx}+f_{lyy}) &
0 \cr
0  &
0  &
f_{l zz} \cr
} \ , 
\label{eq:ffund14}
\end {eqnarray} 
for $l=A$ and $B$. 
$f_{4}$ is the atomic scattering factor of $\rm Mn^{4+}$ ions. 
We emphasize that 
the difference of the normal part of the scattering factor 
$f_{0 A}+f_{0 B}-2f_{0 4}$ 
is of the order of (one electron)/(Mn atom),
which is much smaller than the typical value of the anomalous part at the edge. 
Furthermore, 
the anomalous part of the scattering factor of a $\rm Mn^{4+}$ ion 
is scalar. 
Its value is smaller than that of $\rm Mn^{3+}$ ions 
near the K-edge of a $\rm Mn^{3+}$ ion, because the edge is lower than 
that of a $\rm Mn^{4+}$ ion. \cite{mura1}
Therefore, by utilizing this reflection, 
we can obtain the information of the ferro-component of the scattering factor 
without disturbance of the normal part.
\par
In Fig. 7(a), the 
polarization dependence of the 
scattering intensity at the charge-order reflection is presented. 
The types of the orbital is 
$(\theta_A^{(t)},\theta_B^{(t)})=(4\pi/6,-4\pi/6)$. 
The coordinate in the laboratory system is chosen to be 
$(\hat e_1,\hat e_2,\hat e_3)=
(\hat c, \frac{1}{\sqrt{2}}(\hat a-\hat b), \frac{1}{\sqrt{2}}(\hat a+\hat b))$. 
We neglect the contribution from $f_{0 A}+f_{0B}-2f_{04}$, 
and assume 
that the anomalous part of the scattering factor in a $\rm Mn^{4+}$ ion as 
a scalar, that is, $\Delta f_{4 \alpha \beta}=\Delta f_4 \delta_{\alpha \beta}$, 
and the value of $\Delta f_4$ is 
a half of the maximum value of $  | \Delta f_{l \alpha \alpha}|$ $(l=A,B)$ $(\alpha=x,y,z)$ 
in a $\rm Mn^{3+}$ ion. \cite{mura1}
The explicit formula at $\varphi_A=0$ is given by 
\begin{eqnarray}
\tilde {I} (\varphi,\varphi_A=0)&=& \frac{1}{16}
|(\Delta f_{+xx}+\Delta f_{+yy})\sin^2 \varphi \nonumber \\
&+& 2\Delta f_{+zz}\cos^2 \varphi
-2 \Delta f_4|^2 \ . 
\label{eq:coo}
\end{eqnarray}
In the case of $\varphi=0$ and $\pi$ in Fig. 7(a), 
the electric vector in the incident beam is parallel to the $c$-axis. 
It is worth noting that the polarization dependence in Fig. 7(a) is more remarkable 
in contrast with that shown in Fig. 4(a), because the scalar component of the 
scattering factor is smaller in the present case. 
Furthermore, 
the intensity in the case where the electric vector is parallel to the $ab$-plane 
becomes maximum, that is, the polarization dependence is opposite to the case in Fig. 4(a).  
This is attributed to the fact that the polarization dependence is dominated by 
the anomalous term of $\rm Mn^{3+}$ ions. 
On the other hand, in Fig. 4(a) 
the interference term between the normal and anomalous parts determines the polarization 
dependence. 
As shown in Fig. 7(b), 
the phase of the polarization dependence for $\theta_A^{(t)} > \pi/2$  
is opposite to that for $\theta_A^{(t)} < \pi/2$. 
It is concluded that in the charge ordered state, 
by utilizing the polarization dependence of the charge-order reflection 
the ferro-component of the scattering factor in orbital-AF state is identified easier than 
that in the case of fundamental reflection. 
\section{summary and discussion}
In this paper, we have theoretically investigated the anomalous X-ray scattering 
as a probe to detect the orbital ordering in manganites. 
Through the polarization dependence of the scattering factor, 
we showed how to identify several types of the orbital ordering. 
In particular, 
we paid our attention to 
the method to distinguish the two types of the orbital orderings; 
($3d_{3x^2-r^2},3d_{3y^2-r^2}$) or ($3d_{z^2-x^2},3d_{y^2-z^2}$). 
\par
At the superlattice reflection due to the antiferro-type orbital ordering 
with the condition $\theta_B^{(t)}=- \theta_A^{(t)}$, 
the azimuthal angle dependence is represented by 
a square of the sinusoidal function 
and its intensity depends on types of orbital (Fig. 5). 
We found that it is difficult to 
determine which orbital ordering 
($3d_{3x^2-r^2},3d_{3y^2-r^2}$) or ($3d_{z^2-x^2},3d_{y^2-z^2}$) 
is realized by analyzing the results at the reflection. 
This difficulty is attributed to the fact that 
the antiferro-component 
$\frac{1}{2}(\Delta f_{A \alpha \beta}-\Delta f_{B \alpha \beta})$ 
of the scattering factor is observed at the superlattice reflection. 
Then, we have proposed two kinds of polarization 
analyses where the ferro-component of the scattering factor, i.e., 
$\frac{1}{2}(\Delta f_{A \alpha \beta}+\Delta f_{B \alpha \beta})$, is derived. 
\par
The first one is the polarization analyses at the fundamental reflection (Fig. 4), 
where the scattering factor is represented by sum of the normal part 
and the ferro-component of the anomalous part. 
The interference 
between them gives rise to the polarization dependence 
of the scattering intensity. 
Since the normal part is much larger than the anomalous one,  
the orbital ordering reflects on the modulation in the polarization dependence. 
In order to detect the modulation, 
the reflection plane is required not to include the heavy ions, such as a $\rm La$ ion, 
because the large values of the normal part disturbs to detect the modulation. 
The reflection with the large scattering angle characterized by the large number of $(hkl)$
is also suitable, because the normal part of the scattering 
factor is reduced with increasing the scattering momentum, in contrast with the anomalous part. 
\par
The second proposal to obtain the ferro-component of the scattering factor is 
the analyses at charge-order reflection points in charge ordered phases (Fig. 7). 
At the reflection, 
the normal part is almost canceled out. 
The scattering factor is dominated by the 
ferro-component of the anomalous scattering factor in a $\rm Mn^{3+}$ ion 
and the anomalous part in a $\rm Mn^{4+}$ ion . 
The latter is a scalar and its value is smaller than that of a $\rm Mn^{3+}$ ion near 
the $\rm Mn^{3+}$ K-edge. 
Therefore, by analyzing the polarization dependence of this reflection, 
the ferro-component of the scattering factor is observed.  
It is noted that the basic considerations in the above theoretical proposals 
are independent of details in the numerical microscopic calculation 
and are based on symmetry of 
types of the orbital ordering and the experimental arrangement. 
\par
In conclusion, we have shown that in the anomalous X-ray scattering technique, 
types of the orbital order is able to be identified by 
selecting the adequate reflection point and analyzing the 
polarization dependence of the scattering intensity. 
\par
\medskip
\noindent
ACKNOWLEDEGMENTS
\par
We would like to thank 
Y. Endoh and Y.Murakami for providing the experimental data prior to publication 
and for valuable discussions. We also indebted to W.Koshibae for 
helpful comments. 
This work was supported by Priority Areas Grants from the Ministry of 
Education, Science and Culture of Japan, and CREST (
Core Research for Evolutional Science and Technology Corporation) Japan. 
Part of the numerical calculation was performed in the HITACS-3800/380 supercomputing 
facilities in Institute for Materials Research, Tohoku University. 
\narrowtext

\end{document}